# Application of Reinforcement Learning for 5G Scheduling Parameter Optimization

Ali Asgher Mansoor Habiby[*], Ahamed Mehaboob[**]

[*] Senior 5G/4G/3G/2G Network Optimization Engineer, Vodafone Qatar {aliasgherman@gmail.com}
[**] Senior 5G/4G/3G/2G Radio Access Network Operations Engineer, Vodafone Qatar {ahamedtk@hotmail.com}
{ ali.mansoor@vodafone.com, ahamed.thoppu@vodafone.com }

**Abstract**- *RF Network parametric optimization requires a wealth of experience and knowledge to achieve the optimal balance between coverage, capacity, system efficiency and customer experience from the telecom sites serving the users. With 5G, the complications of Air interface scheduling have increased due to the usage of massive MIMO, beamforming and introduction of higher modulation schemes with varying numerologies. In this work, we tune a machine learning model to 'learn' the best combination of parameters for a given traffic profile using Cross Entropy Method Reinforcement Learning and compare these with RF Subject Matter Expert (SME) recommendations. This work is aimed towards automatic parameter tuning and feature optimization by acting as a Self-Organizing-Network module.*

***Index Terms**- 5G-SON, Reinforcement Learning, Automatic Parameter tuning, 5G Air Interface Scheduling*

## 1. Introduction

Self-Organizing Networks (SON) have been under study [1] for almost a decade along with their suggested use-cases for LTE, however, their wide-ranged usage is yet to be seen [2]. Now with the aggressive rollout of 5G and expectations of a huge number of connected devices [3], improving the efficiency of Network functions is a critical research path. In this work, we focus on Reinforcement learning to automatically suggest and implement parameters that govern the RF scheduling in 5G.

We show that this approach can be used to find the best parameter and feature combinations in an automatic manner that outperforms the recommended settings when suggested by the Subject Matter Experts (SMEs) for 5G Network Optimization.

This method can be utilized as an AI helper system as part of 5G-SON where it would enable the 5G Network Engineers to define their goal in terms of configurable weights between capacity, coverage, and quality without the need for detailed study of the internal working of the parameters and features. This effectively translates into a vendor-implementation independent parameter tuning system.

While this work focuses on ML based optimal parameter and feature selection, previous studies have targeted the selection of waveforms and numerologies using ML as well as utilizing deep learning as a means to optimize the physical layer of 5G. [4], [5].

For our work, we have selected the 5G Downlink scheduler and the related configurable parameters which govern its operation. The scheduler is responsible to ensure the successful delivery of downlink packets received from upper layers (see fig 1 for 5G protocol stack) to the UE with the promised packet delay budget values.

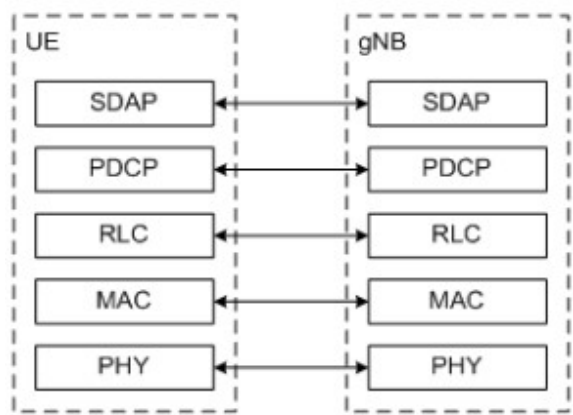

*Figure 1, 5G Protocol Stack. PHY is Layer 1, MAC, RLC, PDCP and SDAP are Layer 2.*

To perform this operation, the scheduler takes into account some sort of scheduling priority like round-robin, proportional fair and other implementations - the scheduler then also calculates the channel conditions of the UE (by means of the UE's reported conditions and ongoing HARQ feedbacks among other parameters) and decides the allocated Modulation Coding Scheme (MCS) along with the number of Resource Blocks (RBs). All this is done with the appropriate consideration of the Block error rate (BLER) due to the inherent demodulation errors during the data reception by the UE. There are further configurable parameters that balance the resource allocation between control channels and data channels (PDCCH and PDSCH) as well as other options of configuration related to Hybrid Automatic Repeat Request (HARQ) and Precoding Matrix Indicator (PMI) processing. Vendor (RAN vendor in our case) based implementation would also add some means of modification of offsets for the internal filters, used with the UE's Channel Quality Indicator (CQI) reports, that are used to avoid high fluctuations and to discourage the use of older CQI reports by the UE which may not be truly reflective of the UE's current state. (fig 2, shows the core inputs and output for a scheduler).

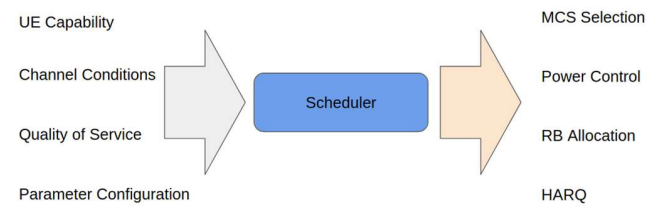

*Figure 2, A brief of inputs and outputs of a scheduler.*

Traditionally, the parameter tuning is performed based on a trial-and-error approach to find out the best value of a single parameter (or a set of parameters in every trial).

$$O = \text{argmax}( K(p) )$$

where,
K is the value of desired KPIs
p is the set of parameters implemented
O is the parameter set which provided the highest gain in KPIs K

In practical terms, it is impossible to evaluate all possible values of the allowed configuration parameters (due to the high number of valid combinations) which can sometimes prevent the 5G Network Engineer from finding the most optimal parameter set.

In order to address the question of finding the best set of parameters automatically, we simplified the traditional approach to a problem of Reinforcement learning.

This is due to the fact that even though all these parameters may serve differing individual purposes, the overall objective is always to get the optimal balance in coverage, capacity and quality for all the UEs served by the 5G cell. (fig 3 shows a sample objective which could be automatically translated into detailed parameters to be implemented to maximize the desired objective)

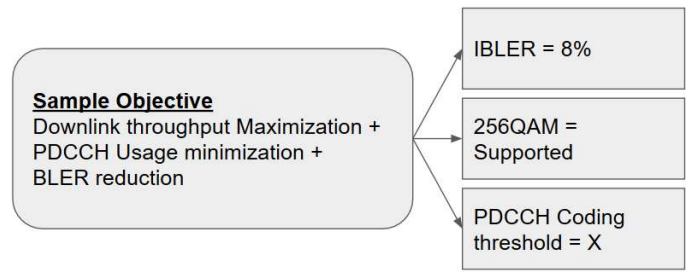

*Figure 3 Definition of an Objective would be automatically translated by the reinforcement algorithm into configuration parameter values*

With this definition in mind, we consider the configurable parameters as *'actions'*, all the other inputs as well as the outputs of the scheduler as the system's *'state'* and then the task becomes to find a policy which maximizes the *'rewards'* which can be defined based on Network operator's preferences on coverage or capacity. These terminologies enable us to translate this issue into a Reinforcement learning problem.

$$Q^\pi(s,a) = \mathrm{E}[R \mid s, a, \pi],$$

Where,
**Q** is the action-value function
**s** is the state which can include the cell throughput, user throughputs, channel conditions averaged and per user, amount of data in the buffer, assigned MCS, CRC errors, etc.
**a** or actions are the configurable parameters which will be changed by the system to maximize reward
**E** is our objective function which defines the KPIs we want to maximize

This work focusses on developing a practical system which can maximize the defined objective to values equal or better than the ones achieved by SME using hand-tuned parameter thresholds. This work also addresses some of the challenges in the implementation of Reinforcement learning to 5G Network parameter tuning. Finally, this work is in line with the key objectives of SON to increase Network efficiency with the use of Machine Learning.

## 2. Methods

### 2.1. The 5G SME

As this whole study is based on a Live 5G site, we also sought help from a 5G Subject Matter Expert (SME) with Planning and Optimization experience on the 5G and 4G Network. This was necessary in order to obtain the values of parameters which are thought of as best-practice and also devise the KPIs which should all be monitored in order to decide whether a combination is performing as per the expectation. Throughout this text, we have mentioned the areas where the inputs were taken from the 5G SME.

### 2.2. Test Methodology

In order to run this experiment within a controlled environment, we selected a 5G cell with no commercial users. We then set up three 5G capable User Equipment (UE) on the 5G cell in excellent coverage, medium coverage, and poor coverage. The coverage thresholds were decided by the SME based on the average expected user experience for these coverage bins. The test devices were configured to automatically generate traffic which included common apps such as Facebook, WhatsApp, YouTube and Ookla SpeedTest. The traffic generation was controlled by a Raspberry Pi connected to each phone using accurately timed scripts so that results are comparable between these devices. The raspberry pi was using Android Debugging Bridge (ADB) to emulate user touch and getting screen readings for the calculation of KPIs such as video loading times, WhatsApp message delivery times, Ookla speed test results etc.

### 2.3. System State Estimation Setup

As stated earlier, we applied a Reinforcement learning algorithm for which we needed to know the system state and the reward functions.

The system state was gathered by the continuous results provided by the test devices which included the Coverage, Throughput, and Latency. Additional system state was captured by online 5G cell trace which captured detailed information including but not limited to DL/UL PDCP bits, DL/UL RLC throughput and bits, DL/UL MAC throughput and bits, DL/UL NACK, DL DTX, DL/UL ACK, UL Interference, DL/UL RBs, PDCCH 2/4/8/16 CCE usage, etc. The system traces were aggregated for every 2000 Transmission Time Intervals (TTIs). In total this generated around 300+ different features.

In order to have a setup to implement different parameter values as recommended by the algorithm, we used a vendor-provided scripting language along with a custom O&M interface to enter commands and implement on the fly.

### 2.4. Reward Function Estimation Setup

The computation of reward-function relied on development of a framework which analysed the traces captured at the Next Generation Node B (gNB) as well as the statistics reported by the test UEs. This framework converted the obtained values from time-binned series into session summaries and included the application of constraints to ensure that each captured sample had approximately similar usage pattern to be comparable. Some of these constraints are defined later in the text.

After this framework was in place, we then asked the 5G SME to provide us the desired characteristics in terms of which KPIs should be improved and in what priorities. The reward function then simply computed a static numeric value based on the SME's assigned priorities to the obtained KPIs.

$$r(s) = \sum_{i=0}^{n} (p_i * k_i(s))$$

Where,
r = Numeric reward for the current state s
i = Index corresponding to the observed KPI
$p_i$ = Weighted priority of the ith KPI where sum of p for $p_0$ to $p_i$ was equal to 1
$k_i$ = the value of the ith observed KPI normalized between 0 to 1 for the state s

For this study, the chosen priorities were,

*Table 1 Chosen priorities of target KPIs for this experiment*

| KPI | Target Priority (%) |
|---|---|
| Downlink MAC Throughput | 22 |
| Downlink RLC Throughput | 29 |
| Downlink ACK Ratio | 28 |
| Uplink ACK Ratio | 15 |
| Downlink Average MCS | 6 |

### 2.5. Action Generation Setup

The actions (or configurable parameters) selected for this work include a mix of features and parameter thresholds. Some of these are shown in Table 1.

These included parameters like IBLER threshold (which control the scheduler's target value of IBLER to converge on to. A higher value may cause the scheduler to always use a very high MCS value causing most blocks to have decoding errors), Features like Adaptive MCS selection (to adaptively switch the MCS selection in order to guarantee a better demodulation performance) and more. In total we selected 10 parameters which included both thresholds as well as Boolean parameters.

*Table 2, Some Parameters which were tuned by the algorithm and their possible values*

| Parameter | Value Range |
|---|---|
| Initial BLER Threshold | 0 to 100% in step of 1% |
| Adaptive MCS Selection | 0 or 1 |
| PMI Enhancement | 0 or 1 |
| Filter of UE MCS value | 0 to 2 in steps of 0.01 |
| Initial Rank Assignment | RANK1 to RANK8 |
| HARQ Enhancement | 0 or 1 |

A simple check reveals that all the possible combinations of these parameters would be in excess of 600K which is too high to be practically tried within a reasonable time frame.

If we consider the RF parameters analogous to hyper-parameters, then it follows that the number of trials would increase exponentially with the number of hyper-parameters similar to a Grid Search [6]. To reduce the number of such combinations we employ a Manual search approach coupled with Randomized parameter set inclusion considering the practical benefits of Random Grid search as shown by [6] should help us converge while the Manual search approach will incorporate domain knowledge specific combinations.

The manual search approach was supported by the 5G SME to select ranges of parameters that seemed most promising. A split of 50/50 was used for Manual recommendations and random value inclusions for the possible parameter values. (our actions)

### 2.6. Environment and Reinforcement Algorithm Setup

Up till this step, the states, actions, and rewards are clearly defined. So we developed a minimal environment similar to OpenAI's gym [7] to provide basic methods to perform full steps with the given actions and return the new state. The environment was responsible for generating the commands to be implemented based on suggested actions and also collecting the state variables and returning them to the analysis framework.

We then implemented a "*Cross-Entropy Method*" (CEM) [8] based Reinforcement learning algorithm which was connected to our designed environment. In order to exclude working on results where no user was scheduled or the scheduled data was not enough, we placed additional constraints that were implemented in the environment.

For a single epoch, we generated 50 sessions where the environment made sure that each of these sessions has at least $X$ seconds where Number of Scheduled Downlink TTIs > 0.8 * 1600 and $Y$ seconds where Number of Scheduled Downlink TTIs is between 0.2*1600 and 0.8*1600. Here 0.8 represents 80% and 1600 is the number of Downlink TTIs which can be present in a 1-second duration for a 5G numerology of 30Khz. Here X and Y were chosen with consultation with 5G SME and later adjusted to complete the full experiment within 3 days.

The CEM in our case used a Multi-Layer Perceptron (MLP) to tune the weights based on the rewards. The total features which were obtained from our environment (inputs to the Neural Network representing the system state) were in excess of 300. The output from the MLP were the parameter values to be implemented (10 total parameters). We initialized the MLP with random weights at the beginning of the experiment.

### 2.7. Baseline Evaluation

Before starting the reinforcement algorithm, we selected the values of all parameters under study based on the 5G SME recommendations which were established upon the recommended values from the RAN vendor. We then ran our scripts on the 5G UEs to collect the system state and convert it into a reward value. We then applied the same constraints listed in the previous paragraph and then selected the highest value of rewards to be considered as a baseline value. The target of this work is to propose parameter values that can be at least equal or better than this baseline reward value.

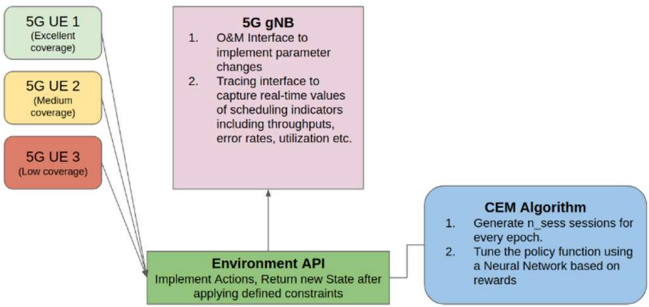

Figure 4, describes the overall system architecture for this work

## 3. Results
### 3.1. CEM Outputs

This setup was under test for 3 days during which approximately 150 epochs were tried. The CEM algorithm kept a real-time track of the baseline value vs. the current rewards. We further tracked the 25th percentile, median, mean, 75th percentile of the rewards in every epoch. These statistics helped us follow the best and worst performances of the model to ensure that a minimum level of customer experience can be ensured. One key difference in the resulting plots is that the baseline value was the maximum reward function whereas the CEM outputs are arranged in their percentile and means. This arrangement was made to discourage some random combinations from CEM achieving higher scores but all other sessions being way lower than the baseline.

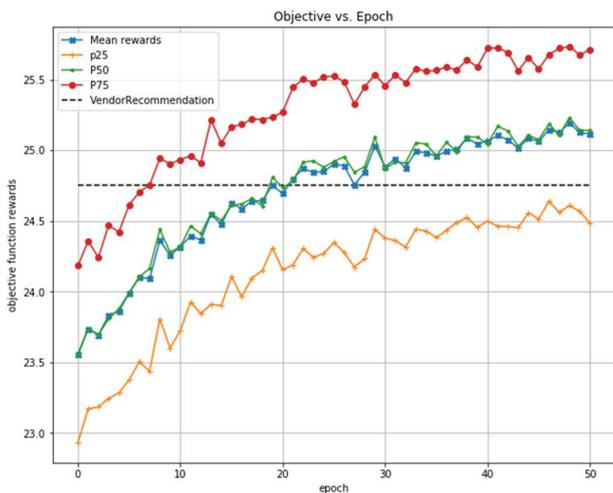

Figure 5, A close-up of the first 50 epochs showing the point where the mean and median rewards crossed the baseline values. This took approximately 10 hours of iterations

As shown in fig 5, the algorithm was able to beat the baseline consistently after 25 epochs. We kept running the algorithm to determine if further reward optimization is possible.

Running further iterations (fig 6) improved the results at a slower pace and by epoch 65 the 25th percentile rewards were almost equal to the baseline.

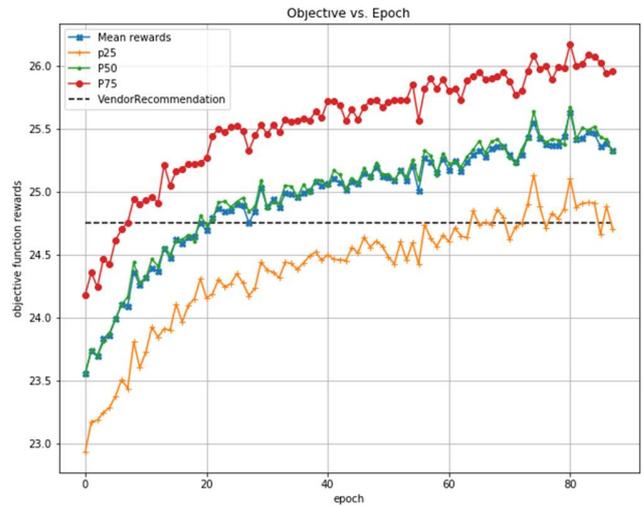

Figure 6, By epoch 65, the 25th percentile rewards were within a reasonable range of the baseline.

With further iterations, the overall rewards did not improve further and by epoch 150 we terminated the experiment. By epoch 140, the algorithm was able to provide 25th percentile rewards which were consistently better than the baseline. (fig 7)

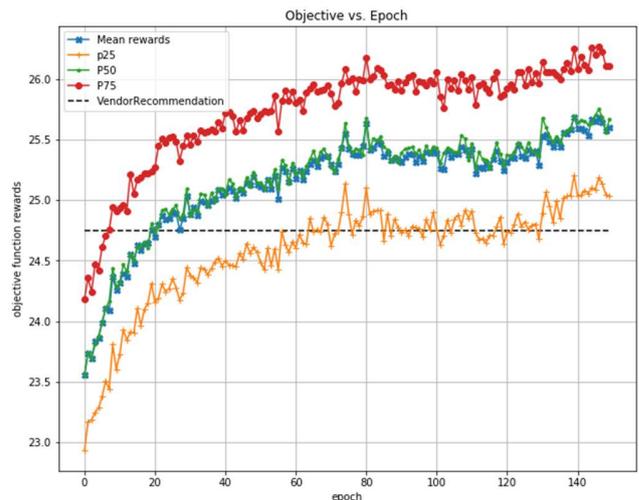

Figure 7, After 150 epochs the system was able to convincingly outperform the baseline values.

Application of Reinforcement Learning for 5G Scheduling Parameter Optimization

### 3.2. Understanding Rewards

In order to understand the values returned by the CEM and the effectiveness of the model, we collected the statistics from traces at each step and compared the system state with the baseline state.

As the reward function was designed to maximize throughputs while minimizing the DTX and NACK ratios, we will focus on these plots to provide a clear understanding of the evolution of these KPIs with every epoch.

Figures 8 to 11 depict the performance of every epoch in terms of KPIs compared to their respective baseline values.

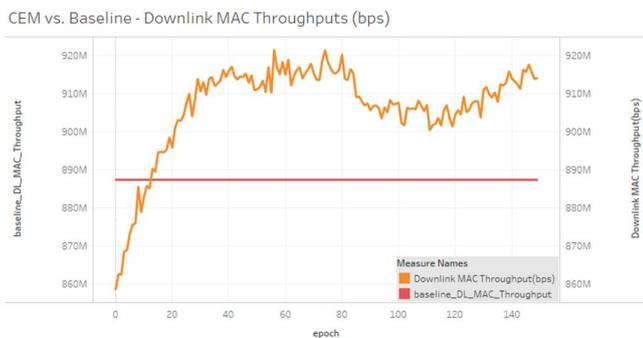

*Figure 8, The algorithm was able to improve the Downlink MAC throughput from the baseline within the first 15 epochs/*

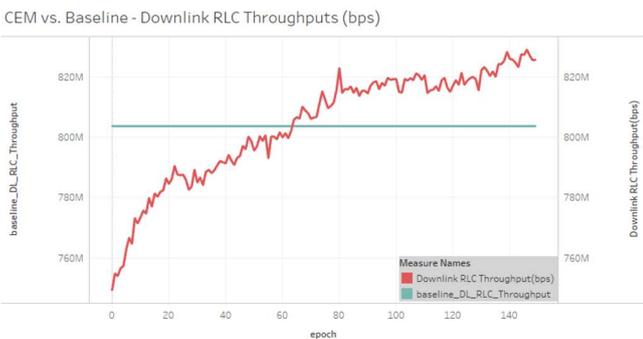

*Figure 9, The algorithm took more than 60 epochs to provide a RLC throughput which was better than the baseline values*

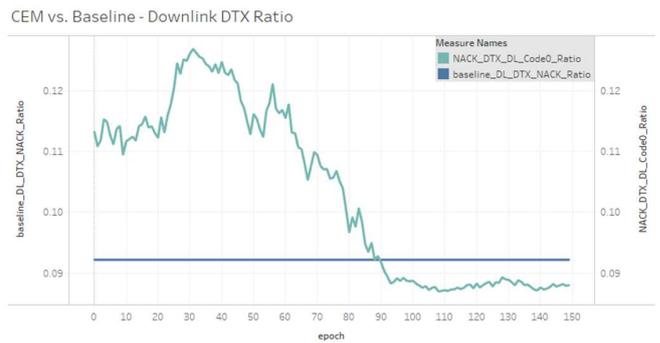

*Figure 10, The algorithm was able to tune the DTX and NACK ratio in the downlink direction better than the baseline by the 90th epoch*

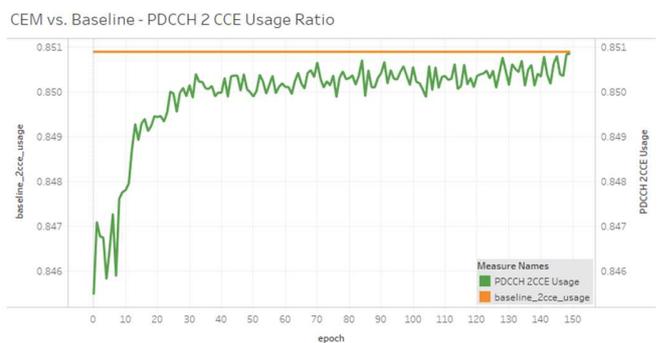

*Figure 11, the algorithm was never able to improve the 2CCE utilization beyond the baseline value. This shows that the assigned priorities in the objective function can have certain KPI values less than the baseline values*

These figures clearly show that a better balance of all KPIs was eventually found by the algorithm with the exception of 2 CCE usage which was given the lowest priority in the objective function definition.

## 4 Conclusion

5G Networks will cater for a huge number of connected devices and will require automated parameter optimization to ensure optimal customer experience with the changing traffic profiles.

A wider support for such SON functionality by the RAN vendors (by means of built-in APIs for feedback loops and parameter configurations) can drastically improve the results from this research area where ML will support the Network Engineers to make better decisions for their customers.